\documentclass[11pt,a4paper]{article}

\usepackage[utf8]{inputenc}
\usepackage[T1]{fontenc}
\usepackage[margin=1in]{geometry}
\usepackage{amsmath}
\usepackage{amssymb}
\usepackage{booktabs}
\usepackage{array}
\usepackage{tabularx}
\usepackage{longtable}
\usepackage{calc}
\usepackage{authblk}
\usepackage{xcolor}
\usepackage{graphicx}
\usepackage{newunicodechar}
\usepackage{listings}
\usepackage[colorlinks=true,linkcolor=blue!40!black,citecolor=blue!40!black,urlcolor=blue!40!black]{hyperref}

\lstdefinestyle{codebox}{
  basicstyle=\small\ttfamily,
  columns=flexible,
  breaklines=true,
  keepspaces=true,
  showstringspaces=false,
  extendedchars=false,
  inputencoding=utf8,
}
\newunicodechar{×}{\ensuremath{\times}}
\newunicodechar{→}{\ensuremath{\rightarrow}}
\newunicodechar{μ}{\ensuremath{\mu}}
\newunicodechar{≈}{\ensuremath{\approx}}
\newunicodechar{≠}{\ensuremath{\neq}}
\newunicodechar{≤}{\ensuremath{\leq}}
\newunicodechar{≥}{\ensuremath{\geq}}

\title{\textbf{From the NYU Ultracomputer to Modern Exascale: A Historical and Architectural Survey of In-Network Computing and Scalable Synchronization}}

        \author[1]{\textbf{Lars Warren Ericson}}
        \affil[1]{Catskills Research Company}
        \affil[1]{\texttt{lars.ericson@catskillsresearch.com}}

        \date{\today}

\begin{document}

        \maketitle

        \begin{center}
          \small
          \textbf{ORCID:} 0000-0001-8299-9361 \\
          \textbf{Primary Category:} cs.DC (Distributed, Parallel, and Cluster Computing) \\
          \textbf{Secondary Category:} cs.AR (Hardware Architecture) \\[0.5em]
          \textbf{Repository:} \url{https://github.com/catskillsresearch/ultracomputer}
        \end{center}

        \begin{abstract}
        This paper presents a historical and technical survey of the hardware architectures, interconnection networks, and synchronization primitives that have shaped massively parallel systems over the past four decades. We examine the design of the NYU Ultracomputer and the IBM Research Parallel Processor Prototype (RP3), focusing on the hardware implementation of the Fetch\&Add primitive in multistage interconnection networks. We contrast these early attempts at fine-grained, shared-memory hardware combining with the distributed-memory architectures of the IBM SP series and the modern in-network computation models found in NVIDIA SHARP and HPE Slingshot. 

We provide a technical analysis of message-passing synchronization, presenting a complete profiling of MPI operation frequencies and detailing the low-level hardware mapping of one-sided RMA atomics to PCIe Atomics and GPU caches. We investigate the software-hardware boundary in modern deep learning, detailing how HIP translation, Triton compilation, and 4-bit quantization (W4A16) execute on modern heterogeneous silicon. 

To evaluate alternative network node designs, we present a historical hardware case study analyzing the feasibility of implementing active combining switches using message-passing Inmos Transputers programmed in Occam. Finally, we contextualize the evolution of concurrent software synchronization by examining Isaac Dimitrovsky’s parallel "group lock" primitive, tracing its downstream echoes in group mutual exclusion (GME) and room synchronization, and reflect on the historical, philosophical divide between American systems engineering and European formal methods.
        \end{abstract}

\hypertarget{introduction}{%
\section{Introduction}\label{introduction}}

The quest for scalable parallel computation has consistently faced two fundamental bottlenecks: physical latency in the interconnect network and memory contention at synchronization hot spots. In the early 1980s, parallel computer design split into two primary paradigms: shared-memory MIMD (Multiple Instruction, Multiple Data) systems utilizing hardware coordination primitives, and distributed-memory message-passing multicomputers.

A central innovation of the shared-memory paradigm was ``in-network computing''---the idea that the routing network itself should participate in the arithmetic reduction of concurrent requests targeting the same memory location. While the physical limits of 1980s Very Large Scale Integration (VLSI) technology ultimately favored the simpler ``dumb-switch, fast-CPU'' message-passing model, the performance bottlenecks of exascale computing and the demands of modern deep learning workloads have driven a major revival of hardware-assisted data reduction.

This paper provides an integrated retrospective on these architectural shifts. Section 2 analyzes the NYU Ultracomputer and the IBM RP3. Section 3 traces the lineage of scalable architectures through the IBM SP series and the Blue Gene paradigm to the modern exascale era. Section 4 examines modern exascale networks, specifically the Dragonfly topology and in-network computing hardware. Section 5 provides a detailed analysis of the software and hardware mechanics of MPI atomics and their fallback workarounds. Section 6 investigates modern deep learning software on heterogeneous hardware, detailing HIP translation and 4-bit quantization mechanics. Section 7 evaluates an alternative historical design point: simulating an active combining switch via Inmos Transputers and Occam. Section 8 investigates Isaac Dimitrovsky's parallel ``group lock'' primitive and its academic descendants. Section 9 reflects on the historical and philosophical divide between American systems engineering and European formal methods.

\begin{center}\rule{0.5\linewidth}{0.5pt}\end{center}

\hypertarget{nyu-ultracomputer-and-ibm-rp3-shared-memory-hardware-combining}{%
\section{NYU Ultracomputer and IBM RP3: Shared-Memory Hardware Combining}\label{nyu-ultracomputer-and-ibm-rp3-shared-memory-hardware-combining}}

\hypertarget{network-geometries-and-node-architecture}{%
\subsection{2.1 Network Geometries and Node Architecture}\label{network-geometries-and-node-architecture}}

The NYU Ultracomputer (designed in 1983) and the IBM RP3 (constructed in 1985) sought to scale shared-memory architectures to hundreds or thousands of processors. In these machines, \(N\) processors were connected to \(N\) memory modules via a multistage interconnection network.

The IBM RP3 architecture utilized two distinct networks operating in parallel: 1. \textbf{Low-Latency Network}: A rectangular SW Banyan network designed for standard memory reads and writes. It provided dual source-sink paths to route around network congestion. 2. \textbf{Combining Network}: A Lawrie Omega network built to route synchronization commands. In the completed 64-processor prototype (the RP3x), this network consisted of six levels of \(2 \times 2\) switches, supplemented by \(4 \times 2\) and \(2 \times 4\) concentrators and deconcentrators to handle routing into the network.

\begin{figure}
\centering
\begin{center}
\includegraphics[width=0.85\textwidth,keepaspectratio]{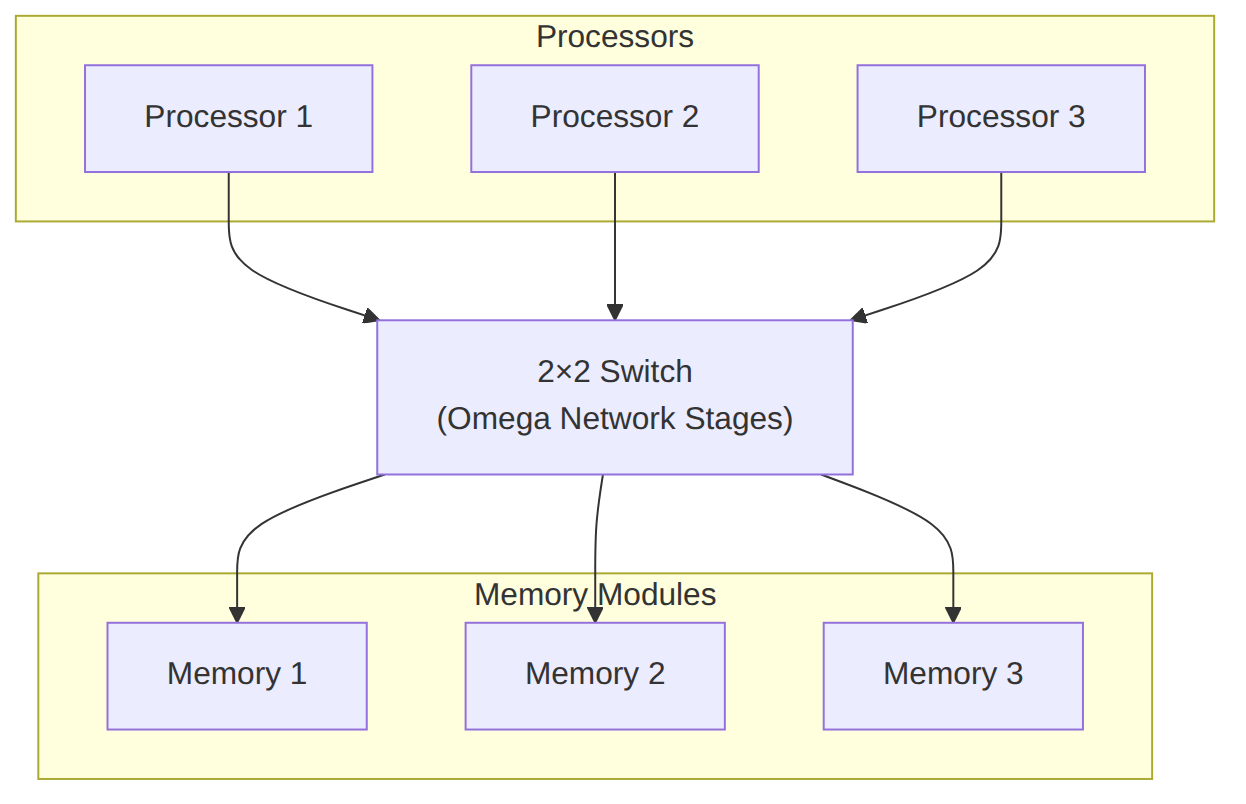}
\end{center}
\caption{Omega Network}
\end{figure}

The physical 64-processor RP3x prototype contained exactly 64 CPUs and a total of 512 MB of RAM. Each Processor-Memory Element (PME) contained 8 Megabytes of primary memory and a 32 KB (later upgraded to 64 KB) cache. A key feature of the RP3 was that the memory could be dynamically partitioned by software to act as local private memory, global shared memory, or a mix of both.

\hypertarget{the-fetchadd-primitive-and-hardware-combining}{%
\subsection{2.2 The Fetch\&Add Primitive and Hardware Combining}\label{the-fetchadd-primitive-and-hardware-combining}}

The defining synchronization mechanism of this architecture was the atomic \textbf{Fetch\&Add (F\&A)} primitive, expressed as: \[\text{F\&A}(X, e)\] where \(X\) is a memory address and \(e\) is an integer increment. This instruction returns the value stored at address \(X\) and concurrently increments \(X\) by \(e\) as an indivisible operation.

To prevent memory serialization when multiple processors target a single shared variable (such as a work-queue counter), the \(2 \times 2\) network switches performed dynamic \textbf{hardware combining}:

\begin{itemize}
\item
  \textbf{Forward Path}: If a switch node received two concurrent Fetch\&Add operations targeting the same memory address---\(\text{F\&A}(X, e_1)\) from \(P_1\) and \(\text{F\&A}(X, e_2)\) from \(P_2\)---the switch's internal ALU intercepted them. The switch added the increments (\(e_1 + e_2\)), stored \(e_1\) in an associative wait buffer, and forwarded a single combined request \(\text{F\&A}(X, e_1 + e_2)\) to the next stage of the network.
\item
  \textbf{At Memory}: The memory controller executed the combined addition as a single atomic operation, returning the original value \(v\).
\item
  \textbf{Backward Path}: When \(v\) returned to the combining switch, the node retrieved \(e_1\) from its buffer. It routed \(v\) back to \(P_1\) and \((v + e_1)\) back to \(P_2\), satisfying both processors with unique, sequentialized return values in a single network round-trip.
\end{itemize}

\hypertarget{the-complexity-wall}{%
\subsection{2.3 The Complexity Wall}\label{the-complexity-wall}}

The VLSI design of the Ultraswitch was spearheaded by a hardware team at NYU's Courant Institute, including Susan Dickey, Richard Kenner (later the primary maintainer of the GNU C Compiler in the early 1990s), Marc Snir, and Jon Solworth. They fabricated custom chips using the MOSIS service under a 1.6-micron double-metal CMOS process.

To make hardware combining work, the switch designed by Dickey and Kenner featured semi-systolic queues and associative wait buffers. When a memory request entered the switch, a hardware comparator checked to see if an identical memory address request was sitting in the queue. If it found a match, the switch's internal ALU mathematically added the values, sent the combined request forward, and held the local state in the associative wait buffer until the memory responded.

The VLSI effort successfully built a fully functional 16-processor, 16-memory module prototype, proving that Jack Schwartz's theory of a ``hot-spot-free'' shared-memory network was physically possible. However, the VLSI implementation hit a harsh physical ceiling:

\begin{itemize}
\item
  \textbf{Pin and Area Limits}: A \(2 \times 2\) switch doing pairwise combining was right at the limit of what could fit on a chip. To scale the Ultracomputer efficiently to thousands of nodes, researchers calculated that the switches would need 3-way or 4-way combining. Scaling the VLSI layout to handle 3-way combining resulted in multi-input adders that took up too much silicon area and severely slowed down the clock speed.
\item
  \textbf{Latency}: The logic required to check the queues, compare addresses, and do the math took several clock cycles. This meant that even when the network was empty and no combining was required, standard memory reads were heavily penalized by the ``smart'' switch's transit latency.
\end{itemize}

Ultimately, the transistor budgets of the 1980s were simply not large enough to make combining switches commercially viable against standard, ``dumb-but-fast'' memory networks.

\begin{center}\rule{0.5\linewidth}{0.5pt}\end{center}

\hypertarget{post-rp3-lineage-the-shift-to-distributed-memory-and-heterogeneity}{%
\section{Post-RP3 Lineage: The Shift to Distributed Memory and Heterogeneity}\label{post-rp3-lineage-the-shift-to-distributed-memory-and-heterogeneity}}

The demise of academic shared-memory supercomputers prompted IBM to pivot toward distributed-memory multicomputers, starting with the SP line.

\hypertarget{the-network-on-the-ibm-sp1-and-sp2}{%
\subsection{3.1 The Network on the IBM SP1 and SP2}\label{the-network-on-the-ibm-sp1-and-sp2}}

The IBM SP1 (introduced in 1993) and SP2 (1994) used an interconnect called the High-Performance Switch (HPS). The HPS was based on the Vulcan switch chip, an \(8 \times 8\) crossbar designed by IBM Research.

The geometry of the HPS was a \textbf{Bidirectional Multistage Interconnection Network (BMIN)}. Unlike the RP3's one-way Omega network where data looped back around, the SP's bidirectional design allowed packets to traverse up the switch hierarchy just far enough to find a common ancestor to the destination node, and then travel back down. It utilized packet-switched, wormhole routing to keep latency low.

Crucially, the SP1 and SP2 did not do Fetch\&Add in the network. They were not shared-memory machines; they were distributed-memory multicomputers. Each node was a self-contained RS/6000 workstation with its own private CPU, RAM, and OS. Because there was no global shared memory, processors could not natively issue a ``Fetch\&Add'' instruction to another node's RAM over the switch. Instead, the SP switch was strictly used for message passing (running protocols like MPI, PVM, and IP). The Vulcan switch chips were simply data routers; they contained no Arithmetic Logic Units (ALUs) and did no mathematical combining of the packets passing through them.

\hypertarget{the-lineage-of-massively-parallel-systems}{%
\subsection{3.2 The Lineage of Massively Parallel Systems}\label{the-lineage-of-massively-parallel-systems}}

The table below traces the architectural lineage and evolution of these systems up to the modern exascale era:

\begin{longtable}[]{@{}
  >{\raggedright\arraybackslash}p{(\columnwidth - 10\tabcolsep) * \real{0.1667}}
  >{\raggedright\arraybackslash}p{(\columnwidth - 10\tabcolsep) * \real{0.1667}}
  >{\raggedright\arraybackslash}p{(\columnwidth - 10\tabcolsep) * \real{0.1667}}
  >{\raggedright\arraybackslash}p{(\columnwidth - 10\tabcolsep) * \real{0.1667}}
  >{\raggedright\arraybackslash}p{(\columnwidth - 10\tabcolsep) * \real{0.1667}}
  >{\raggedright\arraybackslash}p{(\columnwidth - 10\tabcolsep) * \real{0.1667}}@{}}
\toprule\noalign{}
\begin{minipage}[b]{\linewidth}\raggedright
Machine
\end{minipage} & \begin{minipage}[b]{\linewidth}\raggedright
Year
\end{minipage} & \begin{minipage}[b]{\linewidth}\raggedright
Compute Nodes / CPUs
\end{minipage} & \begin{minipage}[b]{\linewidth}\raggedright
Total RAM
\end{minipage} & \begin{minipage}[b]{\linewidth}\raggedright
Interconnect / Topology
\end{minipage} & \begin{minipage}[b]{\linewidth}\raggedright
Processor Architecture
\end{minipage} \\
\midrule\noalign{}
\endhead
\bottomrule\noalign{}
\endlastfoot
\textbf{NYU Ultra3} & 1984 & 16 PEs & Low (MB scale) & Omega Network (VLSI Combining) & Motorola 68010 \\
\textbf{IBM RP3x} & 1988 & 64 PEs & 512 MB & Dual SW Banyan / Omega (Combining) & IBM ROMP (32-bit RISC) \\
\textbf{IBM SP1} & 1993 & Up to 128 nodes & 16 GB max & High-Performance Switch (BMIN) & IBM POWER1 (RS/6000) \\
\textbf{IBM SP2} & 1994 & Up to 512 nodes & 128 GB max & High-Performance Switch 2 (BMIN) & IBM POWER2 (RS/6000) \\
\textbf{IBM RS/6000 SP} & 2000 & 512 SMP nodes (8,192 CPUs) & 6 TB & Colony Switch (SP Switch2) & IBM POWER3-II (Symmetric Multi) \\
\textbf{Blue Gene/L} & 2004 & 65,536 nodes (131,072 PEs) & 32 TB & 3D Torus + Global Tree & PowerPC 440 (Embedded dual-core) \\
\textbf{Blue Gene/Q} & 2012 & 98,304 nodes (1,572,864 PEs) & 1.6 PB & Integrated 5D Torus & PowerPC A2 \\
\textbf{IBM/Nvidia Summit} & 2018 & 4,608 nodes (9,216 CPUs + 27,648 GPUs) & 2.8 PB & Mellanox EDR InfiniBand (Fat-Tree) & IBM POWER9 + Nvidia Volta V100 \\
\textbf{HPE Cray Frontier} & 2022 & 9,408 nodes (9,408 CPUs + 37,632 GPUs) & 4.6 PB & HPE Slingshot-11 (Dragonfly) & AMD EPYC + AMD Instinct MI250X \\
\textbf{HPE Cray El Capitan} & 2024/5 & \textasciitilde11,136 nodes (\textasciitilde44,544 APUs) & \textasciitilde5.6 PB & HPE Slingshot-11 (Dragonfly) & AMD Instinct MI300A APU \\
\end{longtable}

\hypertarget{the-blue-gene-paradigm-and-modern-heterogeneity}{%
\subsection{3.3 The Blue Gene Paradigm and Modern Heterogeneity}\label{the-blue-gene-paradigm-and-modern-heterogeneity}}

The IBM Blue Gene series (2004--2012) departed from heavy, workstation-class symmetric multiprocessing (SMP) nodes, utilizing massive arrays of low-power embedded processors connected via specialized networks. This included a dedicated Global Tree Combining Network engineered specifically to perform mathematical reductions (such as MPI collectives) in hardware.

By the mid-2010s, homogeneous multicore architectures hit a power-efficiency wall. Modern systems resolved this by shifting to heterogeneous nodes where massive GPU accelerators handle the bulk of the floating-point calculations, leaving host CPUs to orchestrate the system. In systems like El Capitan, this boundary is further collapsed through the use of Accelerated Processing Units (APUs) such as the AMD MI300A, which package CPU cores, GPU compute units, and High-Bandwidth Memory (HBM3) onto a single physical substrate with a unified address space.

\begin{center}\rule{0.5\linewidth}{0.5pt}\end{center}

\hypertarget{in-network-computing-in-the-modern-exascale-era}{%
\section{In-Network Computing in the Modern Exascale Era}\label{in-network-computing-in-the-modern-exascale-era}}

Modern exascale networks have reintroduced switch-level combining, but at a different layer of the hardware stack and utilizing different network topologies.

\hypertarget{why-hardware-combining-is-avoided-for-fine-grained-shared-memory}{%
\subsection{4.1 Why Hardware Combining is Avoided for Fine-Grained Shared Memory}\label{why-hardware-combining-is-avoided-for-fine-grained-shared-memory}}

If 1024 CPUs try to lock the same shared memory address simultaneously in a modern shared-memory architecture, they create a ``hot spot.'' However, doing the combining in the transit switches is still avoided for three main reasons: 1. \textbf{The Cache Coherence Nightmare}: Modern CPUs rely heavily on complex cache coherence protocols (MESI, MOESI) and directory-based coherence. If an intermediate network switch catches a memory request, mathematically alters it on the fly, and sends a combined answer back, it bypasses the strict cache coherence rules. Keeping the CPU caches synchronized when switches are secretly doing math on the data is incredibly difficult to engineer reliably. 2. \textbf{The ``Fast Path'' Latency Penalty}: In modern VLSI, transistors are incredibly cheap, so putting an ALU in a switch is easy. The problem is latency. If you add inspection, queuing, and ALU logic into a transit switch, you slow down the routing pipeline. Modern systems prefer to keep the switch as ``dumb and fast'' as physically possible so standard memory reads/writes happen in nanoseconds. 3. \textbf{Better Alternatives Exist}: Instead of doing atomics in the network, modern architectures (like modern x86, ARM, and GPUs) push the atomic logic directly to the Memory Controller or the Shared Last-Level Cache (L3/L4). The transit network just delivers the 1024 requests fast. The memory controller receives them, pipelines them natively in hardware, and rapidly spits the results back out. When combined with smart software algorithms---like MCS locks or hierarchical combining trees, where processors spin on local memory flags rather than hammering a global address---the hot-spot problem is resolved without complicating the network switches.

\hypertarget{topology-dragonfly-vs.-fat-tree}{%
\subsection{4.2 Topology: Dragonfly vs.~Fat-Tree}\label{topology-dragonfly-vs.-fat-tree}}

Historically, large clusters utilized a \textbf{Fat-Tree} (or Clos) topology, which represents a hierarchical ``crossbar of crossbars.'' Fat-Trees require a distinct layer of dedicated spine switches at the top of the hierarchy to route traffic between leaves, necessitating millions of long, expensive optical fiber cables.

Modern exascale systems like El Capitan deploy the \textbf{Dragonfly} topology, which is structured as a \emph{fully connected mesh of fully connected meshes}: 1. \textbf{Intra-Group}: Within a cabinet or local group, switches are connected in an all-to-all mesh using low-cost copper cables. 2. \textbf{Inter-Group}: Each local group has direct optical connections to every other group in the system, forming a global mesh.

\begin{figure}
\centering
\begin{center}
\includegraphics[width=0.85\textwidth,keepaspectratio]{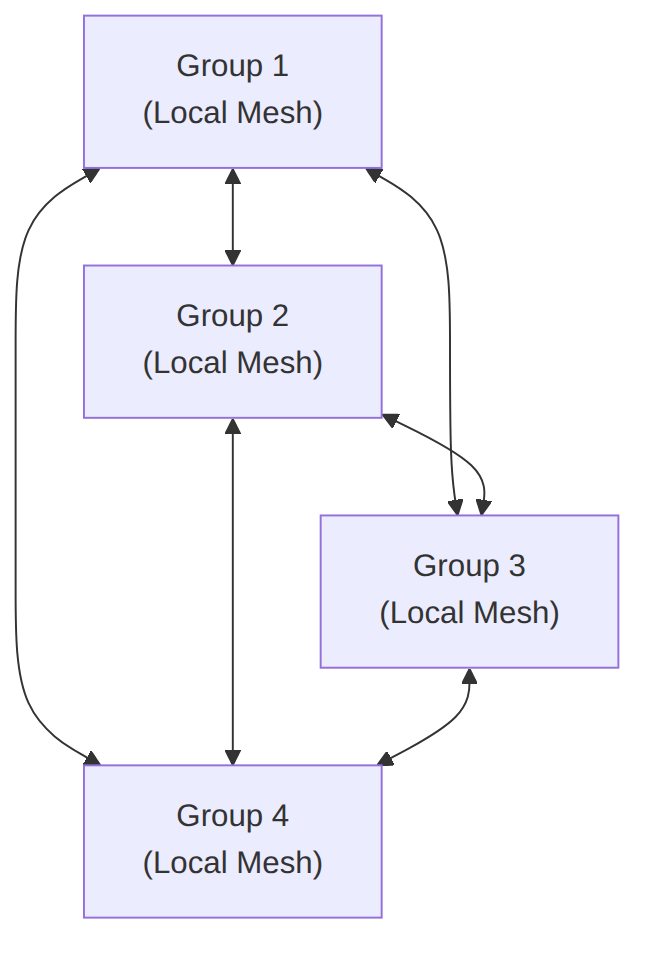}
\end{center}
\caption{Dragonfly Topology}
\end{figure}

The Dragonfly topology eliminates the need for non-compute core switches. Under a standard three-hop routing protocol, any packet can reach its destination in at most three network hops (Local Hop \(\rightarrow\) Global Hop \(\rightarrow\) Local Hop), reducing latency and cabling costs.

\hypertarget{in-network-computing-inc}{%
\subsection{4.3 In-Network Computing (INC)}\label{in-network-computing-inc}}

Instead of implementing fine-grained shared-memory Fetch\&Add operations in transit switches, modern switches target \textbf{message-passing collectives}:

\begin{itemize}
\item
  \textbf{NVIDIA SHARP (Scalable Hierarchical Aggregation and Reduction Protocol)}: Integrated into InfiniBand switch ASICs, SHARP allows the switches to intercept MPI reduction packets (e.g., \texttt{MPI\_Allreduce} used in AI gradient aggregation). The switches execute the reduction arithmetic natively in their internal ALUs at line rate, distributing the aggregated result back to the host nodes and significantly reducing network traffic.
\item
  \textbf{HPE Slingshot (Rosetta Switch ASIC)}: Slingshot switches feature hardware-level arithmetic engines designed to accelerate Partitioned Global Address Space (PGAS) operations and one-sided MPI atomics.
\end{itemize}

This model resolves the 1980s complexity wall. By targeting coarse-grained vector data (such as AI weight gradients) rather than individual, fine-grained memory locking addresses, the latency overhead of switch-level ALUs is amortized over large data payloads.

\hypertarget{software-abstraction-and-scheduling-at-exascale}{%
\subsection{4.4 Software Abstraction and Scheduling at Exascale}\label{software-abstraction-and-scheduling-at-exascale}}

At the node level, El Capitan uses the AMD MI300A APU where CPU cores (Zen 4), GPU compute units (CDNA 3), and 128 GB of shared HBM3 memory are stacked vertically over base I/O dies, connected via 4th Gen AMD Infinity Fabric.

This hardware is managed by performance portability frameworks like \textbf{RAJA} and \textbf{Kokkos}, which translate abstract mathematical loops in C++ into optimized AMD or NVIDIA instructions at compile time.

System resources are managed by advanced batch-job schedulers such as \textbf{Flux} (developed at LLNL), which orchestrate: 1. \textbf{``Hero Runs'' (Monolithic Mode)}: A single massive simulation job is granted exclusive access to all 11,000+ nodes to execute highly integrated multi-physics calculations. 2. \textbf{Standard Operations (Multi-User Partitioned Mode)}: The scheduler mathematically partitions the Dragonfly network, isolating the traffic of multiple users running smaller jobs simultaneously.

\begin{center}\rule{0.5\linewidth}{0.5pt}\end{center}

\hypertarget{deep-dive-the-software-and-hardware-mechanics-of-mpi-atomics}{%
\section{Deep-Dive: The Software and Hardware Mechanics of MPI Atomics}\label{deep-dive-the-software-and-hardware-mechanics-of-mpi-atomics}}

\hypertarget{mpi_allreduce-vs.-fetchadd}{%
\subsection{5.1 MPI\_Allreduce vs.~Fetch\&Add}\label{mpi_allreduce-vs.-fetchadd}}

While both involve combining data across multiple processors, \texttt{MPI\_Allreduce} is not a direct generalization of \texttt{Fetch\&Add}. They serve different purposes, have different semantics, and return different results:

\begin{enumerate}
\def\labelenumi{\arabic{enumi}.}
\item
  \textbf{The Result (Unique Prefix vs.~Global Total)}:

  \begin{itemize}
    \item
    \texttt{Fetch\&Add} gives a unique result to every caller. If Processors A, B, and C simultaneously issue a \texttt{Fetch\&Add(1)} to a memory location that holds 0, the network serializes them. Processor A gets 0, B gets 1, C gets 2, and the memory is updated to 3. This makes it perfect for synchronization---every processor gets a unique ticket to access an array or a queue.
  \item
    \texttt{MPI\_Allreduce} gives the exact same result to every caller. If A, B, and C each submit the value 1 into an \texttt{MPI\_Allreduce(SUM)}, they all get back 3. This is designed for data reduction and computation (such as summing up mathematical gradients), but cannot be used for handing out unique queue tickets.
  \end{itemize}
\item
  \textbf{Synchronization (Asynchronous vs.~Collective)}:

  \begin{itemize}
    \item
    \texttt{Fetch\&Add} is completely asynchronous. A single processor can issue a \texttt{Fetch\&Add} at any time without the participation of other nodes.
  \item
    \texttt{MPI\_Allreduce} is a collective operation. Every single processor in the defined communicator group must actively call the function, or the application will deadlock.
  \end{itemize}
\item
  \textbf{Granularity (Pointers vs.~Vectors)}:

  \begin{itemize}
    \item
    \texttt{Fetch\&Add} operates on a single memory word (e.g., a 32-bit or 64-bit integer) at a specific physical address.
  \item
    \texttt{MPI\_Allreduce} operates on arrays and vectors, performing element-wise reductions across millions of elements simultaneously.
  \end{itemize}
\end{enumerate}

If a developer requires the MPI equivalent of a combined \texttt{Fetch\&Add}, they use \texttt{MPI\_Scan} (Prefix Sum) or \texttt{MPI\_Fetch\_and\_op} (One-Sided RMA Communication).

\hypertarget{mpi-operation-frequency}{%
\subsection{5.2 MPI Operation Frequency}\label{mpi-operation-frequency}}

Continuous profiling at major supercomputing centers using tools like Darshan, mpiP, and IPM reveals the following distribution of MPI operations by call count across general supercomputing workloads:

\begin{longtable}[]{@{}
  >{\raggedright\arraybackslash}p{(\columnwidth - 6\tabcolsep) * \real{0.2500}}
  >{\raggedright\arraybackslash}p{(\columnwidth - 6\tabcolsep) * \real{0.2500}}
  >{\raggedright\arraybackslash}p{(\columnwidth - 6\tabcolsep) * \real{0.2500}}
  >{\raggedright\arraybackslash}p{(\columnwidth - 6\tabcolsep) * \real{0.2500}}@{}}
\toprule\noalign{}
\begin{minipage}[b]{\linewidth}\raggedright
Rank
\end{minipage} & \begin{minipage}[b]{\linewidth}\raggedright
MPI Operation Family
\end{minipage} & \begin{minipage}[b]{\linewidth}\raggedright
Approx. \% of Total Calls
\end{minipage} & \begin{minipage}[b]{\linewidth}\raggedright
Primary Use Case
\end{minipage} \\
\midrule\noalign{}
\endhead
\bottomrule\noalign{}
\endlastfoot
\textbf{1} & \texttt{MPI\_Isend} / \texttt{MPI\_Irecv} & 40\% -- 50\% & Non-blocking point-to-point data exchange (Halo exchanges in grid solvers) \\
\textbf{2} & \texttt{MPI\_Wait} / \texttt{MPI\_Waitall} & 25\% -- 35\% & Resolving non-blocking sends and receives \\
\textbf{3} & \texttt{MPI\_Send} / \texttt{MPI\_Recv} & 10\% -- 15\% & Traditional blocking point-to-point communication \\
\textbf{4} & \texttt{MPI\_Allreduce} & 3\% -- 8\% & Global data reduction (Summing errors, dot products) \\
\textbf{5} & \texttt{MPI\_Bcast} & 1\% -- 3\% & Broadcasting configurations or parameters \\
\textbf{6} & \texttt{MPI\_Test} / \texttt{MPI\_Probe} & 1\% -- 2\% & Checking for incoming messages asynchronously \\
\textbf{7} & Other Collectives (\texttt{Gather}, \texttt{Scatter}) & \textless{} 1\% & Distributing arrays across nodes \\
\textbf{8} & Environment (\texttt{Init}, \texttt{Comm\_rank}, etc.) & \textless{} 0.5\% & Setup and teardown of the application \\
\textbf{9} & RMA (\texttt{Put}, \texttt{Get}, \texttt{Fetch\_and\_op}) & \textless{} 0.1\% & One-sided communication, atomic counters, load balancing \\
\end{longtable}

\emph{Note: Even though \texttt{MPI\_Allreduce} has a lower call count percentage than Send/Recv, it is often the function where the application spends the highest percentage of its time due to network synchronization.}

\hypertarget{workarounds-for-mpi_fetch_and_op}{%
\subsection{5.3 Workarounds for MPI\_Fetch\_and\_op}\label{workarounds-for-mpi_fetch_and_op}}

If \texttt{MPI\_Fetch\_and\_op} were removed from the MPI standard, developers would have to resort to four major workarounds, ranging in performance and complexity:

\begin{enumerate}
\def\labelenumi{\arabic{enumi}.}
\item
  \textbf{The Direct Alternative (\texttt{MPI\_Get\_accumulate})}:

  \begin{itemize}
    \item
    \emph{Mechanism}: \texttt{MPI\_Get\_accumulate} is the vector-based equivalent of \texttt{Fetch\_and\_op}. To replace \texttt{Fetch\_and\_op}, developers can call \texttt{Get\_accumulate} and set the array length to 1.
  \item
    \emph{Overhead}: It requires a slightly heavier function signature and can incur a tiny software overhead inside some MPI implementations, but is functionally identical.
  \end{itemize}
\item
  \textbf{The CAS Loop (\texttt{MPI\_Compare\_and\_swap})}:

  \begin{itemize}
    \item
    \emph{Mechanism}: To safely add 1 to a remote variable, a processor first reads the remote value (e.g., 5), locally calculates the new value (6), and then sends an \texttt{MPI\_Compare\_and\_swap} saying: \emph{``If the remote value is still 5, change it to 6.''} If another processor changed the value in the meantime, the CAS fails, and the loop retries.
  \item
    \emph{Overhead}: CAS loops over high-latency networks degrade performance rapidly under heavy contention, saturating the network with failed retries.
  \end{itemize}
\item
  \textbf{Explicit Window Locks (\texttt{MPI\_Win\_lock} / \texttt{MPI\_Win\_unlock})}:

  \begin{itemize}
    \item
    \emph{Mechanism}: This involves locking the remote memory space, executing a standard \texttt{MPI\_Get} to read the value, adding 1 locally, executing \texttt{MPI\_Put} to write it back, and unlocking the memory.
  \item
    \emph{Overhead}: This bypasses hardware-level acceleration, forcing the system to perform heavy software synchronization.
  \end{itemize}
\item
  \textbf{The Manager-Worker Model (MPI-1 Fallback)}:

  \begin{itemize}
    \item
    \emph{Mechanism}: One dedicated processor (a ``Manager'' rank) acts as the counter. Every other processor sends a standard \texttt{MPI\_Send} message requesting a ticket. The Manager rank runs a loop, receives the message, increments a local variable, and replies with \texttt{MPI\_Send}.
  \item
    \emph{Overhead}: The Manager rank becomes a severe serialization bottleneck, and an entire CPU core is lost simply to host a variable.
  \end{itemize}
\end{enumerate}

\hypertarget{under-the-hood-hardware-mapping-of-atomics}{%
\subsection{5.4 Under-the-Hood Hardware Mapping of Atomics}\label{under-the-hood-hardware-mapping-of-atomics}}

When \texttt{MPI\_Fetch\_and\_op} is called over a top-tier interconnect, the MPI library maps it directly to native hardware silicon:

\begin{itemize}
\item
  \textbf{InfiniBand (NVIDIA/Mellanox ConnectX)}: The MPI library translates the call into the InfiniBand network instruction \texttt{IBV\_WR\_ATOMIC\_FETCH\_AND\_ADD}. The NIC shoots a network packet across the switch. Upon arrival at the destination NIC, the target CPU is not interrupted. The destination NIC uses PCIe Atomics to send a hardware signal directly to the host computer's memory controller. The memory controller's physical ALU does the math, writes the new value to RAM, and sends the old value back to the NIC, which routes it back to the origin node.
\item
  \textbf{NVLink (GPU to GPU)}: An atomic operation from one GPU to another GPU in the same chassis maps directly to NVLink silicon. The sending GPU executes an atomic instruction routed directly over NVLink cables to the receiving GPU's L2 Cache and Memory Controller partition. Dedicated atomic arithmetic units sitting next to the L2 cache lines execute the math in hardware, leaving the receiving GPU's streaming multiprocessors uninterrupted.
\item
  \textbf{HPE Slingshot (Rosetta Switch \& Cassini NIC)}: The Cassini NICs plug directly into PCIe Gen 5 lanes. The Rosetta switches and Cassini NICs have dedicated, integrated arithmetic engines that manage atomic locks and update host memory, ensuring microsecond latency.
\end{itemize}

If run on standard hardware (like basic Gigabit Ethernet) that lacks hardware support, the MPI implementation (e.g., OpenMPI) falls back to a software \textbf{Active Message} workaround: 1. The origin process sends a standard network message: \emph{``Please add 1 to memory address 0x1234.''} 2. A hidden background thread (async progress thread) on the target machine receives the message and wakes the target CPU. 3. The target CPU takes out a local software lock, reads the memory, adds 1, saves it, and sends a standard network message back.

This software fallback takes 10 to 50+ microseconds (compared to 1 to 3 microseconds for hardware execution) and introduces operating system jitter that degrades parallel scaling.

\begin{center}\rule{0.5\linewidth}{0.5pt}\end{center}

\hypertarget{modern-deep-learning-software-on-heterogeneous-hardware}{%
\section{Modern Deep Learning Software on Heterogeneous Hardware}\label{modern-deep-learning-software-on-heterogeneous-hardware}}

\hypertarget{the-hip-translation-layer}{%
\subsection{6.1 The HIP Translation Layer}\label{the-hip-translation-layer}}

In PyTorch, the target device name for AMD hardware remains \texttt{"cuda"}. To make porting AI models as frictionless as possible, AMD built a translation layer called \textbf{HIP (Heterogeneous-compute Interface for Portability)}. When PyTorch is compiled for the ROCm platform, PyTorch intercepts any command targeting \texttt{.cuda()} or \texttt{device="cuda"} and automatically routes it through the HIP compiler directly to the AMD silicon. Consequently, \texttt{torch.cuda.is\_available()} returns \texttt{True} on an AMD-based supercomputer like El Capitan.

\hypertarget{the-cudnn-moat-and-triton-compilation}{%
\subsection{6.2 The cuDNN Moat and Triton Compilation}\label{the-cudnn-moat-and-triton-compilation}}

Historically, NVIDIA's proprietary \texttt{cuBLAS} (for matrix math) and \texttt{cuDNN} (for deep neural networks) libraries formed an effective software barrier. AMD bypassed this by developing open-source drop-in replacements:

\begin{itemize}
\item
  \textbf{rocBLAS}: AMD's direct replacement for \texttt{cuBLAS}.
\item
  \textbf{MIOpen}: AMD's direct replacement for \texttt{cuDNN}.
\end{itemize}

When running PyTorch on AMD, the library is compiled to call \texttt{MIOpen} instead of \texttt{cuDNN}. Standard operations---such as convolutions in a CNN or linear layers---map directly to AMD's hardware AI Accelerators (utilizing Wave Matrix Multiply-Accumulate or \textbf{WMMA} instructions) using \texttt{rocBLAS} via PyTorch's Automatic Mixed Precision (\texttt{torch.amp.autocast}).

This software layer is further abstracted by \textbf{Triton}, an open-source programming language developed by OpenAI. In PyTorch 2.0, \texttt{torch.compile()} looks at the Python code and dynamically generates custom Triton code on the fly. The AMD ROCm backend for Triton then compiles this intermediate language down to raw AMD machine code.

\hypertarget{remaining-barriers-for-non-nvidia-hardware}{%
\subsection{6.3 Remaining Barriers for Non-NVIDIA Hardware}\label{remaining-barriers-for-non-nvidia-hardware}}

While the software gap has closed significantly, NVIDIA maintains a software advantage in several areas: 1. \textbf{Day-1 GitHub Repositories}: Bleeding-edge models often bypass standard PyTorch APIs and hardcode custom CUDA C++ extensions, requiring developers to manually port them to ROCm/HIP. 2. \textbf{FlashAttention Optimization}: FlashAttention kernels are typically written in highly optimized NVIDIA CUDA assembly first, with the ROCm versions lagging behind in features and peak optimization. 3. \textbf{Quantization Ecosystem}: Many high-performance quantization libraries (such as \texttt{bitsandbytes} for 4-bit/8-bit operations) were historically CUDA-only, though ROCm support has been increasingly integrated.

\hypertarget{the-mechanics-of-4-bit-quantization-w4a16}{%
\subsection{6.4 The Mechanics of 4-bit Quantization (W4A16)}\label{the-mechanics-of-4-bit-quantization-w4a16}}

When running large language models in 4-bit quantization (such as GGUF, AWQ, or GPTQ), the GPU hardware does not actually execute 4-bit arithmetic. Instead, it uses \textbf{W4A16 (Weights are 4-bit, Activations are 16-bit)} quantization as a memory bandwidth optimization:

\begin{enumerate}
\def\labelenumi{\arabic{enumi}.}
\item
  \textbf{Storage}: The model's neural network weights are compressed and stored in VRAM as 4-bit values.
\item
  \textbf{Transfer}: The GPU pulls these 4-bit values across the memory bus. Because they are compressed, this requires 75\% less bandwidth than 16-bit weights.
\item
  \textbf{Dequantization}: Once the 4-bit values arrive inside the GPU's registers and ultra-fast SRAM, an unpacking kernel (written in HIP C++ or Triton) dequantizes them back into 16-bit floats (FP16 or BF16).
\item
  \textbf{Math}: The GPU's hardware AI Accelerators perform standard 16-bit matrix multiplications.
\end{enumerate}

Because 4-bit quantization is primarily a strategy to overcome memory capacity and bandwidth limitations, GPUs with larger physical VRAM buffers hold a significant advantage. For example, the AMD Radeon RX 7900 XTX features 24 GB of VRAM compared to the 16 GB found on the similarly priced NVIDIA RTX 4080, allowing larger quantized models to fit entirely within the fast physical memory buffer.

\begin{center}\rule{0.5\linewidth}{0.5pt}\end{center}

\hypertarget{emulating-the-combining-switch-transputers-and-occam}{%
\section{Emulating the Combining Switch: Transputers and Occam}\label{emulating-the-combining-switch-transputers-and-occam}}

As an alternative to custom VLSI layouts, early parallel designs could theoretically be prototyped using off-the-shelf programmable parallel hardware. The \textbf{Inmos Transputer} (introduced in the mid-1980s) was uniquely suited for this task. It natively implemented Communicating Sequential Processes (CSP) concurrency via the \textbf{Occam} language and featured four high-speed, bidirectional serial links.

\hypertarget{structural-concept-of-a-transputer-based-omega-node}{%
\subsection{7.1 Structural Concept of a Transputer-Based Omega Node}\label{structural-concept-of-a-transputer-based-omega-node}}

To build a \(2 \times 2\) combining switch, a single Transputer (such as the T414) can be mapped to route between two processor-side links and two memory-side links:

\begin{itemize}
\item
  \textbf{Link 0 (In/Out A)}: Connected to Processor-Side Port A
\item
  \textbf{Link 1 (In/Out B)}: Connected to Processor-Side Port B
\item
  \textbf{Link 2 (In/Out 0)}: Connected to Memory-Side Port 0
\item
  \textbf{Link 3 (In/Out 1)}: Connected to Memory-Side Port 1
\end{itemize}

Because Transputer links are serial and bidirectional, the physical parallel buses of custom VLSI switches are replaced with simple twisted-pair wires, eliminating pin-count limitations.

\hypertarget{occam-implementation-of-a-combining-switch-node}{%
\subsection{7.2 Occam Implementation of a Combining Switch Node}\label{occam-implementation-of-a-combining-switch-node}}

Below is a stylized Occam implementation of a \(2 \times 2\) combining switch node. It demonstrates how incoming Fetch\&Add requests are evaluated for collision, combined, and subsequently de-combined on their return path:

\begin{verbatim}
-- Protocol for Fetch&Add packets
-- Format: [Memory Address, Increment Value]
PROTOCOL packet IS INT; INT:

-- The 2x2 Ultraswitch Node
PROC ultraswitch(CHAN OF packet inA, inB, out0, out1,
                 CHAN OF packet mem.resp0, mem.resp1, backA, backB)

  INT addrA, valA, addrB, valB:
  INT resp.addr, resp.val:
  
  -- Associative Wait Buffer for 1 combined state
  INT wait.addr, wait.valA:
  BOOL is.combined:

  SEQ
    is.combined := FALSE
    
    -- Main switch loop
    WHILE TRUE
      ALT
        -- ==========================================
        -- BACKWARD PATH: Receive response from Memory
        -- ==========================================
        mem.resp0 ? resp.addr; resp.val
          IF
            -- Did we combine this address earlier?
            (is.combined AND (resp.addr = wait.addr))
              SEQ
                PAR
                  -- Return original value (v) to Sender A
                  backA ! resp.addr; resp.val
                  
                  -- Return prefixed sum (v + valA) to Sender B
                  backB ! resp.addr; (resp.val + wait.valA)
                
                -- Clear associative buffer
                is.combined := FALSE
                
            -- Standard, uncombined memory return path
            TRUE
              backA ! resp.addr; resp.val

        -- ==========================================
        -- FORWARD PATH: Intercept and Combine
        -- ==========================================
        inA ? addrA; valA
          -- PRI ALT peeks at Channel B to capture concurrent collisions
          PRI ALT
            inB ? addrB; valB
              IF
                -- Collision Detected: identical target address
                addrA = addrB
                  SEQ
                    -- Store State of Sender A in Wait Buffer
                    wait.addr := addrA
                    wait.valA := valA
                    is.combined := TRUE
                    
                    -- Forward single combined request to Memory
                    out0 ! addrA; (valA + valB) 

                -- No collision: route packets sequentially
                TRUE
                  SEQ
                    out0 ! addrA; valA
                    out0 ! addrB; valB
            
            -- No immediate packet on B, forward A directly
            TRUE & SKIP
              out0 ! addrA; valA
:
\end{verbatim}

\hypertarget{cost-and-performance-trade-offs}{%
\subsection{7.3 Cost and Performance Trade-offs}\label{cost-and-performance-trade-offs}}

Evaluating the cost and performance of a 64-processor Omega network implemented with Transputer switches (requiring 192 T414 chips) reveals a classic hardware-software trade-off:

\begin{enumerate}
\def\labelenumi{\arabic{enumi}.}
\item
  \textbf{Physical Complexity \& Cost}: At a mid-1980s pricing of \(\sim \$400\) per T414 chip, the raw silicon cost of the switch fabric would be approximately \(\$76,800\), plus board integration costs. While expensive, this design eliminates the massive multi-layer printed circuit board routing and custom VLSI masking fees associated with dedicated hardware switches.
\item
  \textbf{The Latency Penalty}: Custom VLSI switches processed parallel memory requests in nanoseconds. In contrast, the Transputer routes data over serial links. Transferring a 64-bit packet (32-bit address + 32-bit value) over a 10 Mbps serial link takes \(\approx 6.4\ \mu\text{s}\) per hop. Across a 6-stage Omega network, the round-trip latency would exceed \(100\ \mu\text{s}\). This delay is orders of magnitude slower than contemporary DRAM access times (\(\approx 100\text{ ns}\)), showing that while a Transputer-based combining network is programmatically elegant, it is physically impractical for fine-grained shared memory.
\end{enumerate}

\begin{center}\rule{0.5\linewidth}{0.5pt}\end{center}

\hypertarget{advanced-synchronization-primitives-isaac-dimitrovskys-parallel-group-lock}{%
\section{Advanced Synchronization Primitives: Isaac Dimitrovsky's Parallel Group Lock}\label{advanced-synchronization-primitives-isaac-dimitrovskys-parallel-group-lock}}

As parallel architectures evolved, academic research turned toward software primitives that could leverage Fetch\&Add to construct bottleneck-free synchronization barriers.

\hypertarget{origin-of-the-group-lock}{%
\subsection{8.1 Origin of the Group Lock}\label{origin-of-the-group-lock}}

In his 1988 PhD thesis at New York University, \emph{``ZLISP---a portable parallel LISP environment''} (advised by Malcolm C. Harrison), Isaac Aaron Dimitrovsky introduced the \textbf{group lock} primitive. First proposed as an NYU Ultracomputer Technical Report in November 1986 (\emph{``A Group Lock Algorithm with Applications''}) and later published in the \emph{Journal of Parallel and Distributed Computing} (1991), the group lock was designed to allow dynamic groups of processes to coordinate without centralized serialization bottlenecks.

Unlike traditional mutual exclusion locks (which serialize processes) or static barriers (which require a fixed number of participating processors), the group lock enables:

\begin{itemize}
\item
  \textbf{Dynamic Membership}: Processes can join and leave coordinating groups dynamically.
\item
  \textbf{Scalable Queue/Stack Operations}: By splitting a group lock into two distinct parts separated by an internal synchronization phase, a program can decouple push (enqueue) operations from pop (dequeue) operations.
\item
  \textbf{Fetch\&Add Optimization}: The group lock can be implemented entirely using wait-free Fetch\&Add and Fetch\&Increment instructions. It bypasses the need for super-step counters by cycling state variables, which minimizes network traffic on shared-memory architectures.
\end{itemize}

\hypertarget{downstream-echoes-and-academic-impact}{%
\subsection{8.2 Downstream Echoes and Academic Impact}\label{downstream-echoes-and-academic-impact}}

Over the last forty years, Dimitrovsky's group lock has influenced parallel coordination algorithms:

\begin{figure}
\centering
\begin{center}
\includegraphics[width=0.85\textwidth,keepaspectratio]{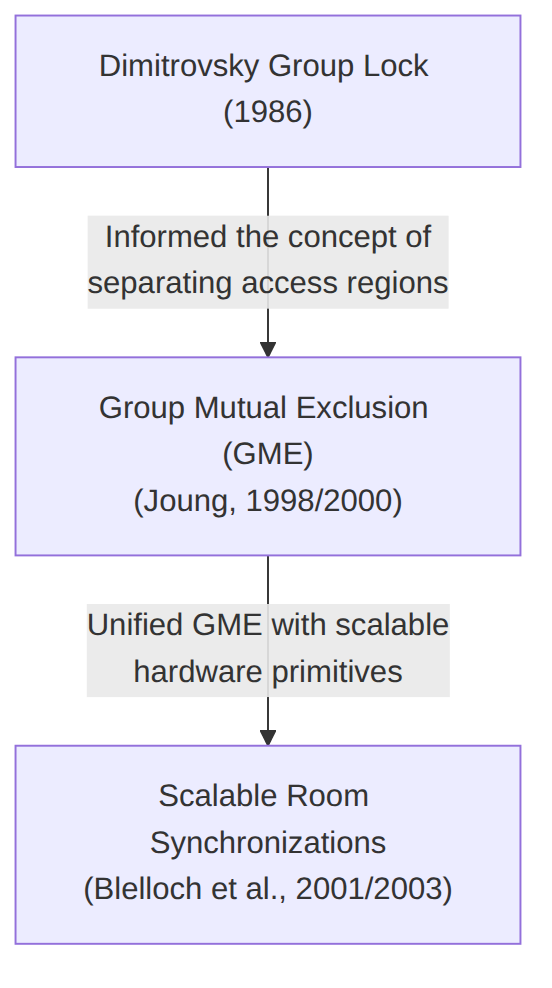}
\end{center}
\caption{Group Lock Lineage}
\end{figure}

\begin{enumerate}
\def\labelenumi{\arabic{enumi}.}
\item
  \textbf{Precursor to Group Mutual Exclusion (GME)}:\\
  Dimitrovsky's group lock is recognized as an early, concrete implementation of the \emph{Group Mutual Exclusion} problem---later formalized by Joung in 1998--2000. In GME, processes request access to different ``sessions'' (or groups) such that processes accessing the same session can enter concurrently, while processes targeting different sessions are excluded.
\item
  \textbf{Impact on ``Room'' Synchronizations}:\\
  In their seminal work \emph{``Scalable Room Synchronizations''} (2001/2003), Guy E. Blelloch, Perry Cheng, and Phillip B. Gibbons compared their ``rooms'' protocol directly to Dimitrovsky's group lock: \textgreater{} \emph{``Dimitrovsky suggests a similar technique for implementing stacks and queues. Instead of using multiple rooms, he uses a single `group lock.' By splitting the group lock into two parts with a synchronization in the middle he is able to separate the pushes from the pops.''}

  They noted that while the group lock lacked a formal proof of linearizability and was less general than their multi-room synchronization model, it pioneered the technique of using split-lock synchronization to implement highly concurrent parallel queues and stacks without bottlenecks.
\item
  \textbf{Integration into Parallel OS Kernels}:\\
  Dimitrovsky's work was integrated into the research and development of the \textbf{Symunix-2} operating system for the NYU Ultracomputer. The group lock and concurrent parallel hash table algorithms he designed allowed the operating system to manage memory allocation and task queues without incurring serial bottlenecks at the kernel level.
\end{enumerate}

\begin{center}\rule{0.5\linewidth}{0.5pt}\end{center}

\hypertarget{historical-and-philosophical-case-study-systems-pragmatism-vs.-formal-purity}{%
\section{Historical and Philosophical Case Study: Systems Pragmatism vs.~Formal Purity}\label{historical-and-philosophical-case-study-systems-pragmatism-vs.-formal-purity}}

The development of parallel systems in the late 1970s and early 1980s was not merely a series of engineering efforts; it was characterized by a deep philosophical division between \textbf{American Systems Engineering} and \textbf{European Formal Methods}.

This division is illustrated by a historical encounter at Carnegie Mellon University (CMU) between 1980 and 1983. Tony Hoare, the creator of Communicating Sequential Processes (CSP), was presenting his formal algebraic framework. During the session, an undergraduate researcher working on the DARPA Distributed Sensor Network (DSN) project under Rick Rashid---using early asynchronous interprocess communication (IPC) protocols and the precursors to the Mach operating system---asked a pragmatic question:

\begin{quote}
\emph{``How does CSP work in a physical, distributed environment with variable latency and potential node failures?''}
\end{quote}

Hoare responded with a classic formalist dismissal, asking: \emph{``How many additions are there in \(7\)?''}

To a formalist, CSP represented a pristine, closed mathematical algebra. The physical realities of distributed systems---such as clock drift, network routing latency, dropped packets, and partial node failures---were treated as implementation details that lay beneath the mathematics. If a physical network dropped a packet, the network was broken, not the mathematical model. To Hoare, physical parallelism on separate CPUs was mathematically reducible to the arbitrary interleaving of sequential events on a single processor.

Conversely, the systems engineering perspective of the CMU group was rooted in physical pragmatism. They were building operating systems (such as Accent and Mach) that had to operate on physical hardware, where asynchronous IPC and fault tolerance were the defining design constraints.

This division ultimately shaped the trajectory of parallel computing:

\begin{itemize}
\item
  \textbf{Brittle Formalism}: When David May and Inmos attempted to implement Hoare's pure, synchronous CSP model in the Transputer and the Occam language, they quickly encountered the constraints of the physical world. To make Occam viable on actual hardware, they had to introduce pragmatic, non-CSP extensions, such as the \texttt{ALT} construct equipped with physical timeout timers. The perfectly synchronous rendezvous model of CSP proved too brittle to scale across the messy, heterogeneous distributed networks that emerged in the 1990s.
\item
  \textbf{Pragmatic Success}: Meanwhile, the pragmatic, asynchronous message-passing models developed at CMU (such as Mach's asynchronous IPC) became the foundation of modern operating systems, microservices, and mobile platforms, including the Mach-derived kernels that power modern macOS and iOS systems.
\end{itemize}

This case study demonstrates that while formal mathematical frameworks are valuable for proving correctness within a closed system, scalable parallel architectures must ultimately prioritize physical constraints---such as latency, power density, and asynchronous communication---to achieve long-term viability.

\begin{center}\rule{0.5\linewidth}{0.5pt}\end{center}

\hypertarget{conclusion}{%
\section{Conclusion}\label{conclusion}}

The historical trajectory of parallel system design reveals a recurring cycle: hardware architectures alternate between shared-memory structures and message-passing designs, but the fundamental mathematical challenges of synchronization remain constant. The hardware-combining network proposed by the NYU Ultracomputer and realized in the IBM RP3 was initially defeated by the economic and physical constraints of 1980s VLSI silicon. However, the core concept of in-network computation has been validated at exascale, where switches in systems like El Capitan handle data reduction to bypass cache-coherency bottlenecks.

Similarly, the programming paradigms developed during the early days of parallel computing continue to resonate. The CSP model, exemplified by Occam and the Inmos Transputer, foreshadowed modern microservice architectures. Concurrently, software synchronization primitives such as Isaac Dimitrovsky's parallel group lock paved the way for scalable, linearizable data structures and room synchronization protocols. Modern parallel computing systems continue to build on these historical foundations.

\begin{center}\rule{0.5\linewidth}{0.5pt}\end{center}

\hypertarget{acknowledgements}{%
\section{Acknowledgements}\label{acknowledgements}}

The human authors retain sole responsibility for the historical claims, architectural descriptions, citations, and conclusions in this survey. Following standard publisher practice (e.g., COPE guidance on authorship and AI tools {[}COPE24{]}), \textbf{no large language model is listed as a co-author}---authorship implies accountability that automated systems cannot bear.

We gratefully acknowledge assistance from the following tools:

\textbf{Cursor} ({[}Cur25{]}): agent-assisted editing in the Cursor IDE, including models routed through Cursor's \textbf{Auto} agent mode (which may invoke Composer-family and other backend models depending on task). These agents helped draft and revise survey prose, convert ASCII figures to Mermaid diagrams, and format Occam and mathematical notation. Generated text was treated as provisional until verified against primary sources and reviewed by the human authors.

\textbf{Google Gemini 3.5 Flash} ({[}Gem25{]}): independent technical briefs on combining-network hardware, Dragonfly topology, and modern in-network reduction (SHARP and Slingshot). Those briefs informed subsequent human-directed revisions; we did not adopt every recommendation verbatim without cross-checking against the cited literature.

All factual claims, diagram semantics, code excerpts, and final prose were reviewed and owned by the human authors. Intellectual property in this note rests with the authors under the project's stated license.

\begin{center}\rule{0.5\linewidth}{0.5pt}\end{center}

\hypertarget{references}{%
\section{References}\label{references}}

\begin{enumerate}
\def\labelenumi{\arabic{enumi}.}
\item
  G. E. Blelloch, P. Cheng, and P. B. Gibbons, ``Scalable Room Synchronizations,'' \emph{Theory of Computing Systems}, vol.~36, no. 5, pp.~327--359, 2003.
\item
  S. Dickey, R. Kenner, and M. Snir, ``An Implementation of a Combining Network for the NYU Ultracomputer,'' \emph{Ultracomputer Note \#93}, Courant Institute, NYU, 1986.
\item
  I. A. Dimitrovsky, ``A Group Lock Algorithm with Applications,'' \emph{Technical Report}, Courant Institute, New York University, Nov.~1986.
\item
  I. A. Dimitrovsky, ``ZLISP---a portable parallel LISP environment,'' Ph.D.~dissertation, Dept. Comput. Sci., New York University, 1988.
\item
  I. A. Dimitrovsky, ``The group lock and its applications,'' \emph{Journal of Parallel and Distributed Computing}, vol.~11, no. 4, pp.~291--302, Apr.~1991.
\item
  J. Edler, \emph{Practical Structures for Parallel Operating Systems}, Ph.D.~dissertation, Dept. Comput. Sci., New York University, 1995.
\item
  A. Gottlieb, R. Grishman, C. P. Kruskal, K. P. McAuliffe, L. Rudolph, and M. Snir, ``The NYU Ultracomputer---Designing an MIMD Shared Memory Parallel Computer,'' \emph{IEEE Transactions on Computers}, vol.~C-32, no. 2, pp.~175--189, Feb.~1983.
\item
  Y. Joung, ``Asynchronous group mutual exclusion,'' \emph{Distributed Computing}, vol.~13, no. 4, pp.~189--206, 2000.
\item
  G. F. Pfister, W. C. Brantley, D. A. George, S. L. Harvey, W. J. Kleinfelder, K. P. McAuliffe, E. A. Melton, V. A. Norton, and J. Weiss, ``The IBM Research Parallel Processor Prototype (RP3): Introduction and Architecture,'' in \emph{Proceedings of the International Conference on Parallel Processing}, 1985, pp.~764--771.
\item
  Committee on Publication Ethics (COPE). (2024). Authorship and AI tools: COPE position statement. https://publicationethics.org/guidance/cope-position/authorship-and-ai-tools
\item
  Anysphere, Inc.~Cursor: AI-native code editor and agent environment. https://cursor.com (accessed 2025).
\item
  Google DeepMind. (2025). Gemini model family (including Flash). Technical documentation and model cards. https://ai.google.dev/gemini-api/docs/models
\end{enumerate}

\end{document}